\documentclass{PoS}
\def\EE{entanglement entropy }
\def\TE{thermal entropy }

\def\beq{\begin{equation}}
\def\enq{\end{equation}}

\def\Tr{{\rm Tr}\,}
\usepackage{braket}
\title{Approaching conformality with the Tensor Renormalization Group method }
\ShortTitle{Near-conformality with the TRG}

\author{\speaker{Yannick Meurice}\\
        Department of Physics and Astronomy, The University of Iowa, Iowa City, Iowa 52242, USA 
        E-mail: \email{yannick-meurice@uiowa.edu}}
\author{Li-Ping Yang\\      Department of Physics,Chongqing University, Chongqing 401331, China \\
E-mail: \email{liping2012@cqu.edu.cn}}
       \author{Judah Unmuth-Yockey
 \\
        Department of Physics and Astronomy, The University of Iowa, Iowa City, Iowa 52242, USA 
        E-mail: \email{judah-unmuth-yockey@uiowa.edu}}

\author{Yuzhi Liu\\ Department of Physics, University of Colorado, Boulder, Colorado 80309, USA\\E-mail:\email{Yuzhi.Liu@colorado.edu}}
 \author{James Osborn\\ALCF, Argonne National Laboratory, Argonne, IL 60439, USA\\
       E-mail: \email{osborn@alcf.anl.gov}}
\author{Z. Y. Xie\\ Institute of Physics, Chinese Academy of Sciences, P.O. Box 603, Beijing 100190, China\\E-mail:\email{qingtaoxie@ruc.edu.cn} }
\author{Haiyuan Zou\\ Department of Physics and Astronomy, University of Pittsburgh, Pittsburgh, PA, USA\\ E-mail: \email{haz57@pitt.edu}}

\abstract{ We discuss the Tensor Renormalization Group (TRG)  method for the $O(2)$ model with a chemical potential 
in 1+1 dimensions with emphasis
on near gapless/conformal situations. We emphasize the role played by the late Leo Kadanoff in the 
development of this theoretical framework. 
We describe the entanglement entropy in the superfluid phase (see arXiv:1507.01471 for details). 
We present recent progress on optimized truncation 
methods. 
}

\FullConference{The 33rd International Symposium on Lattice Field Theory\\
		14 -18 July 2015\\
		Kobe International Conference Center, Kobe, Japan*}

\begin{document}

\section{Introduction}

The idea of considering the average spin in cells of variable sizes, originally put forward by Leo Kadanoff \cite{Kadanoff:1966wm}, has played a crucial role in the development of the renormalization group ideas. Leo Kadanoff's untimely death a few weeks ago should make us reflect on the fact that theoretical innovations can take a long time to be practically realized. 
Despite its great visual appeal, the blockspinning procedure is not easy to implement numerically. It typically involves approximations that are difficult to improve. It took several decades to develop numerical methods where blocking can be achieved with controllable errors. 
Ultimately, the question of continuum limits of lattice models should be addressed directly with the blocking procedure.

The Tensor Renormalization Group (TRG) method combined with higher order SVD were used to construct accurate blocking 
methods to deal with the thermodynamics  of the Ising model \cite{PhysRevB.86.045139}. The TRG provides 
an {\it exact} block spinning procedure which separates 
the degrees of freedom inside the block (integrated
over) from those kept to communicate with the
neighboring blocks.  For the Ising model, two state approximations provide accurate exponents in 2D ($\nu$ = 0.99) 
but less accurate exponents  in 3D ($\nu$ = 0.74) \cite{prb87}. We realized that adding a few more states, does not immediately improve the results. Similar findings were reported 
by Leo Kadanoff and collaborators \cite{efratirmp}. This has remained a puzzle and Leo was convinced that its resolution may involve understanding approximate hierarchical symmetries. 
Hopefully younger generations will help make progress in this direction. 

The TRG formulation  for lattice models with compact gauge or spin fields can be facilitated by the use of character expansions. 
This was used to derive {\it exact} blocking formulas for the $O(2)$ and $O(3)$ models, principal chiral models, Abelian gauge theories and $SU(2)$ gauge theories  \cite{Exactblocking13prd}. 
The truncation procedure relies on the diagonalization of positive matrices and there seems to be no sign problems \cite{prd89} associated with the method. 
Fermions can be included without sign problem, for instance for the Schwinger model \cite{Shimizu:2014uva} or the Gross-Neveu model  \cite{Takeda:2014vwa}. 
The TRG method allows to smoothly connect the relativistic formulation where space and time are treated on equal footing, with the Hamiltonian formalism. 
This feature has been exploited to propose quantum simulators for the $O(2)$ model \cite{PhysRevA.90.063603} and the Abelian Higgs model \cite{PhysRevD.92.076003}.
The transfer matrix formalism developed in Ref. \cite{PhysRevA.90.063603} allows us to introduce finite temperature or external field effects. 
It would be interesting to connect the TRG formalism with the Matrix Product States calculations of finite temperature \cite{Saito:2014bda} or external electric field effects \cite{Buyens:2015dkc}. 
TRG methods are also used in the context of spin foams \cite{Dittrich:2014mxa}. The blocking program has evolved in many directions and is alive.

In these proceedings, we discuss TRG calculations for the $O(2)$ model with a chemical potential $\mu$. More details can be found in Ref. \cite{Yang:2015rra} and related studies  in Refs.   \cite{Bruckmann:2014sla,Bruckmann:2015sua,Kawauchi:2015heu}. 
In Ref. \cite{Yang:2015rra}, we compared the TRG calculation of the particle density in the superfluid phase with a sampling algorithm \cite{Banerjee:2010kc} developed using a classical version of the worm algorithm \cite{PhysRevLett.87.160601} and found that 
the distributions obtained with the two methods agree well.  A transfer matrix formulation of the TRG method briefly explained below can be used to calculate the thermal entropy and
the entanglement entropy.
As we increase $\mu$ to go across the superfluid phase between the first two Mott insulating phases, and for
a sufficiently large temporal size, we see an interesting fine structure: the average particle number and the winding number of most of the world lines in the Euclidean time direction increase by one unit at a time.  At each step, the thermal entropy develops a peak and the entanglement entropy increases until we reach half-filling, and then decreases in a way that approximately mirrors the ascent. This suggests an approximate fermionic picture which can be qualitatively
understood from the approximate correspondence with the spin-1/2 XY quantum chain when $\mu$ is large enough to justify a two-state approximation. We also report progress in optimization following earlier attempts \cite{Meurice:2014tca,Unmuth-Yockey:2014afa}.

\section{TRG formulation of the $O(2)$ model with a chemical potential $\mu$}
\begin{figure}[h]
  \includegraphics[width=0.23\textwidth,angle=0]{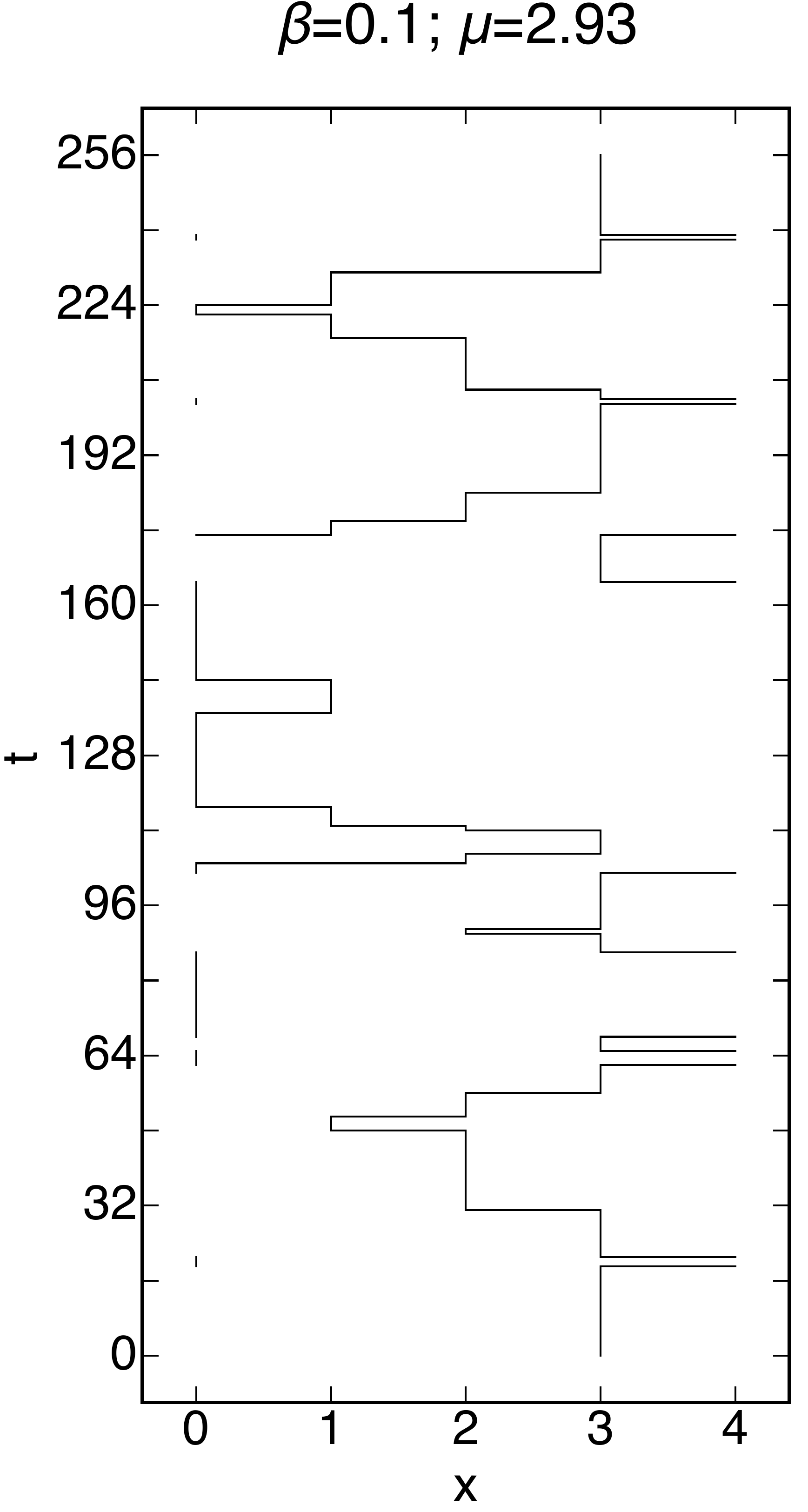}
   \includegraphics[width=0.23\textwidth,angle=0]{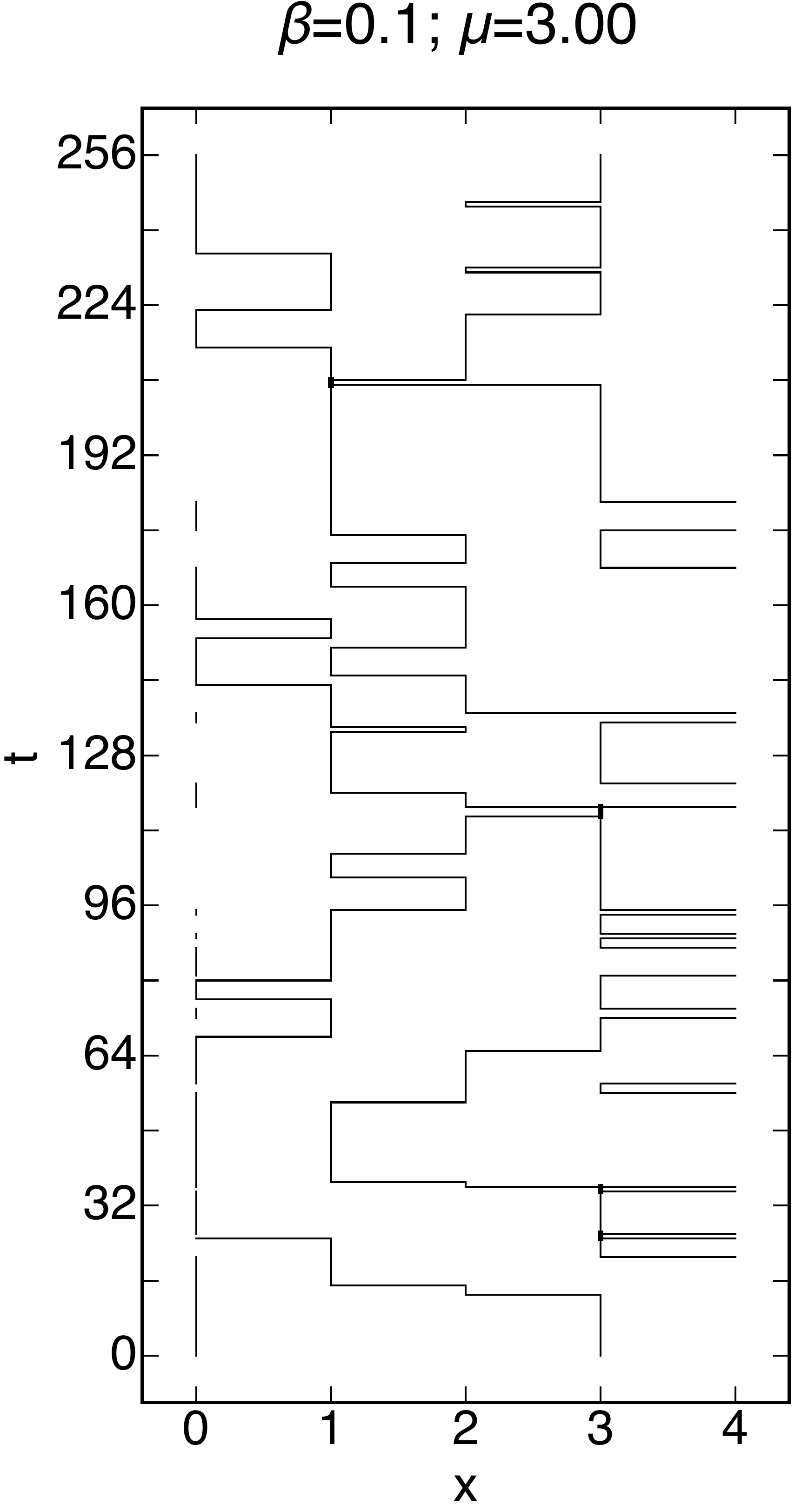}
   \includegraphics[width=0.23\textwidth,angle=0]{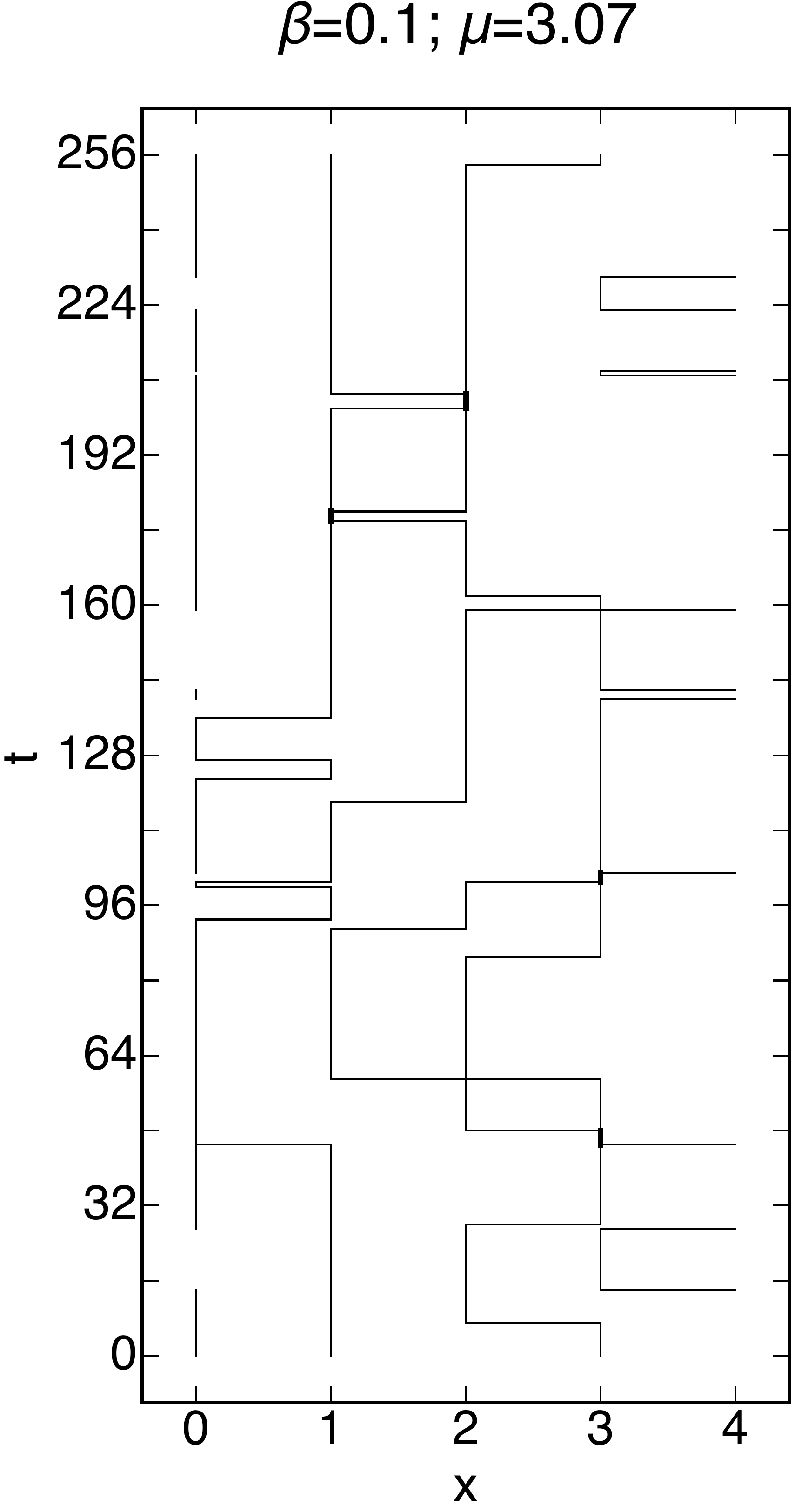}
   \includegraphics[width=0.23\textwidth,angle=0]{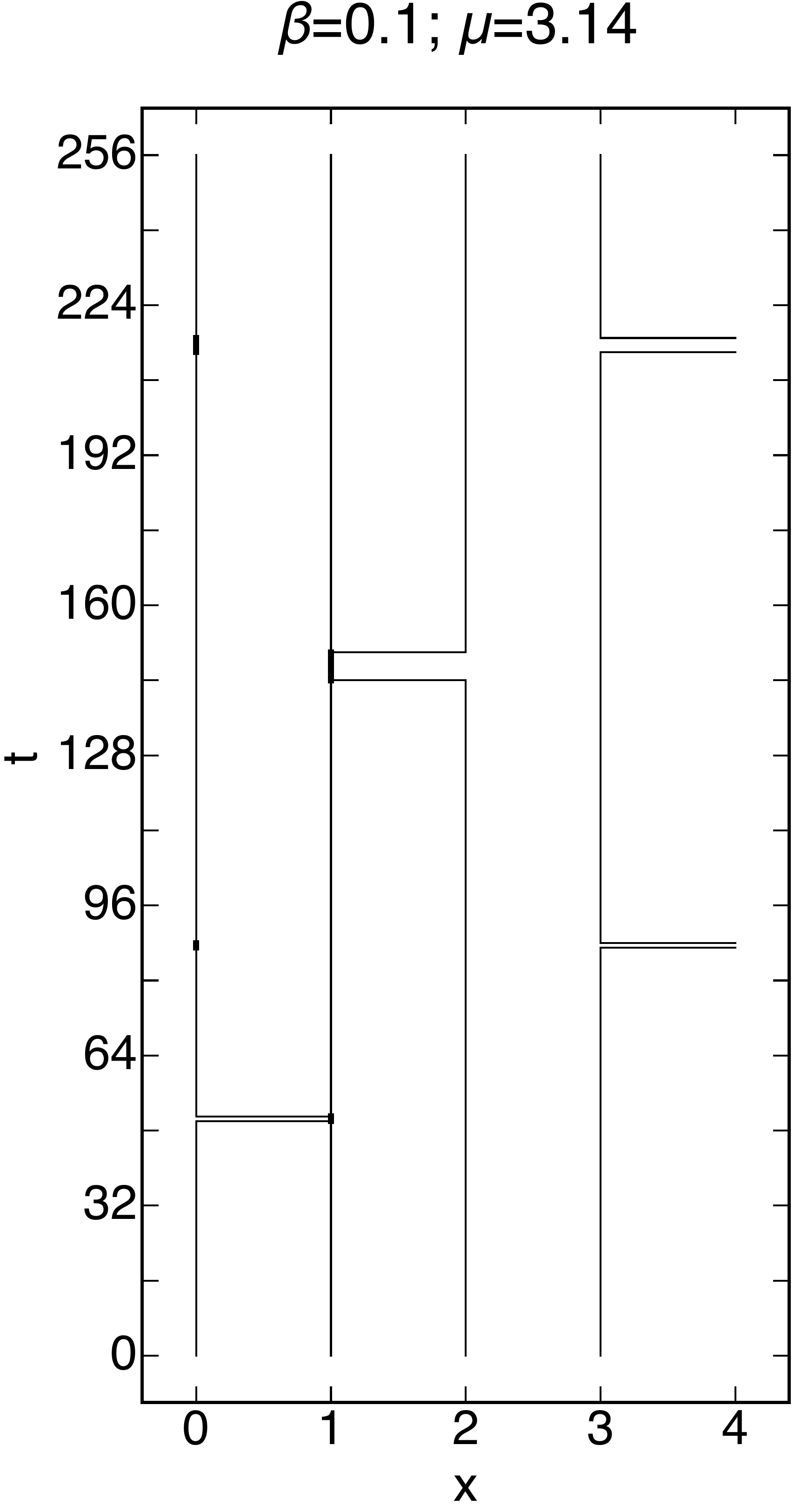}
\caption{ \label{fig:allworm} Worm configurations for  $\mu$ =2.93, 3.00, 3.07 and 3.14  for $\beta=0.1$, $L_x=4$ and $L_t=256$. Almost all the $|n|$' s are 0 or 1. Between most time slices we have $n$ time links carrying a current 1 and $L_x-n$ time links carrying no current. In rare occasions, lines merge or cross (see Ref. \cite{Yang:2015rra} for details).}
\end{figure}

The partition function for the model on a $L_t\times L_x$ lattice can be written as 
        \beq
            Z = \int{\prod_{(x,t)}{\frac{d\theta_{(x,t)}}{2\pi}} {\rm e}^{-S}} \ ,
            \label{eq:bessel}
        \enq
with the action
\beq
      S=  -\beta_{\hat{t}}\sum\limits_{(x,t)} \cos(\theta_{(x,t+1)} - \theta_{(x,t)}-i\mu)-\beta_{\hat{x}}\sum\limits_{(x,t)} \cos(\theta_{(x+1,t)} - \theta_{(x,t)}).
      \enq
The action is complex, but a positive representation is obtained using character expansions \cite{RevModPhys.52.453}: 
\beq
Z =\sum_{\{n\} }\prod_{(x,t)} I_{n_{(x,t),\hat{x}}}(\beta_{\hat{x}})I_{n_{(x,t),\hat{t}}}(\beta_{\hat{t}}){e^{\mu n_{(x,t),\hat{t}}}}\\ 
 \times\delta_{n_{(x-1,t),\hat{x}}+n_{(x,t-1),\hat{t}},n_{(x,t),\hat{x}}+n_{(x,t),\hat{t}}} \ . 
\label{eq:Z_worm}
\enq
The Kronecker delta function reflects the particle number conservation which can be exploited to construct 
a worm algorithm \cite{Banerjee:2010kc}  and sample the allowed worm configurations (see Fig. \ref{fig:allworm}).
The $n$ variables associated with the links are sometimes mistakenly called dual variables. In two dimensions, the dual variables \cite{RevModPhys.52.453} are associated with the plaquettes of the dual lattice. In continuum language,  the plaquette field $\phi$ is used to get a divergenceless vector $n^\mu=\epsilon^{\mu\nu}\partial _\nu \phi$.

The partition function can be written as $Z=\Tr\mathbb{T}^{L_t},$ with $\mathbb{T}$ the transfer matrix. 
Following  Ref. \cite{PhysRevA.90.063603}, 
the matrix elements of $\mathbb{T}$ can be expressed as a product of tensors associated with the sites of a time slice (fixed $t$) and traced over the space indices 
$$\mathbb{T}_{(n_1,n_2,\dots n_{{L_x}})(n_1',n_2'\dots n_{L_x}')}=\sum_{\tilde{n}_{1}\tilde{n}_{2}\dots \tilde{n}_{L_x}} T^{(1,t)}_{\tilde{n}_{L_x}\tilde{n}_{1}n_1 n_1'}T^{(2,t)}_{\tilde{n}_{1}\tilde{n}_{2}n_2n_2'\dots }\dots  T^{(L_x,t)}_{\tilde{n}_{L_{x-1}}\tilde{n}_{L_x}n_{L_x}n_{L_x}' }$$
with
$$T^{(x,t)}_{\tilde{n}_{x-1}\tilde{n}_x n_x n_x'} =  \sqrt{I_{n_{x}}(\beta_{\hat{t}})I_{n'_x}(\beta_{\hat{t}})I_{\tilde{n}_{x-1}}(\beta_{\hat{x}})I_{\tilde{n}_x}(\beta_{\hat{x}}){\rm e}^{(\mu(n_x+n_x'))}} \delta_{\tilde{n}_{x-1}+n_x,\tilde{n}_x+n_x'}$$
The transfer matrix can be constructed using a hierarchical blocking procedure when $L_x=2^q$. We can express the space contraction of two tensors with 4 indices as 
a new tensor with 4 indices (this involves a projection) and repeat $q$ times (see Refs. \cite{PhysRevA.90.063603,Yang:2015rra}). 
 The phase diagram can be constructed by considering the changes in the particle density as $\mu$ increases. In the Mott insulating (gapped) phase, the density stays constant, while in the superfluid phase (SF) it increases with $\mu$. 
 As we reach the SF phase, the gap disappears and we expect conformal symmetry to be present. In addition, the thermal and entanglement entropy develop a non-trivial behavior described in the next section. 
 The phase diagram in the $(\beta,\mu)$ plane is shown below in Fig. \ref{fig:phased2s} together with the thermal entropy in a small region of the phase diagram (see also lower right corner of Fig. \ref{fig:net}).
 \begin{figure}[hh]
\advance\leftskip 1cm
\includegraphics[width=2.7in]{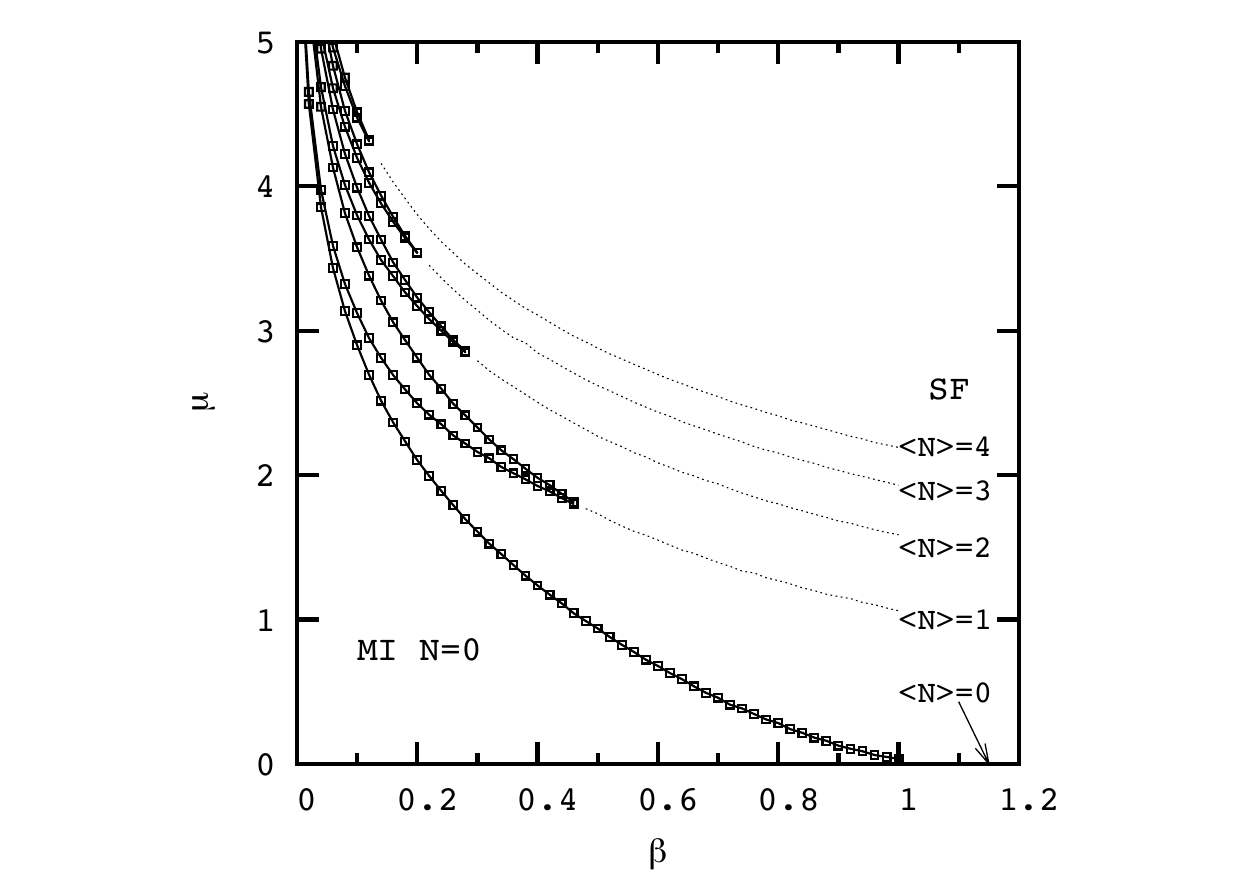}
 \includegraphics[width=1.9in]{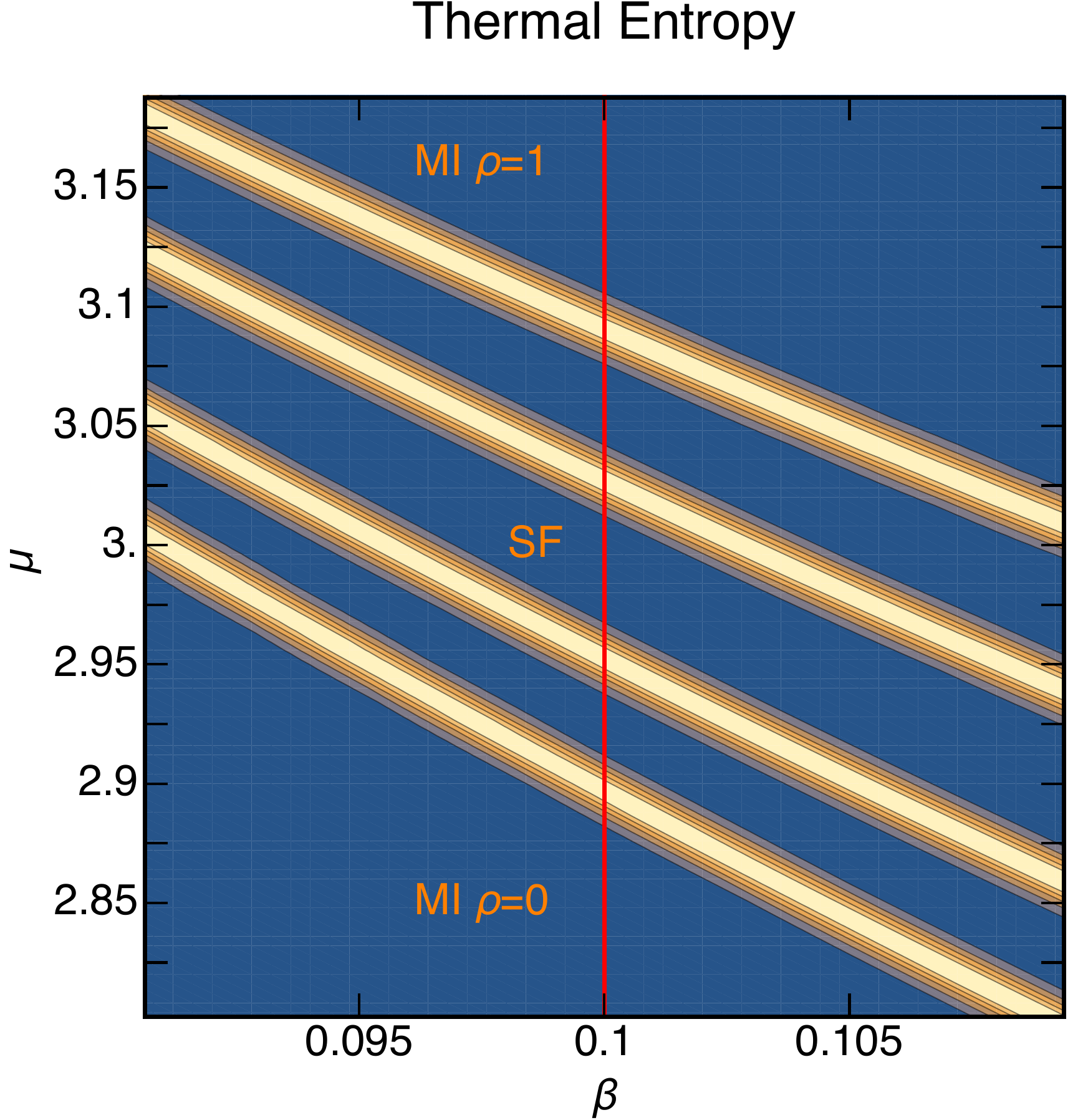}
 \caption{\label{fig:phased2s} Left: Mott Insulating ``tongues" with particle density 0, 1, 2 ... from bottom to top. Right: thermal entropy in a small region of the $\beta-\mu$ plane on a $4\times128$ lattice. 
 The blue regions are close to zero and the yellow ochre regions peak near $\ln 2$. The MI phase 
 with $\rho$=0 is below the lowest light band, the MI phase with $\rho$=1 is above the highest light band and there is a single SF phase in between the two MI phases.}
\end{figure}
\section{Thermal entropy and entanglement entropy}
\begin{figure}[h]
\begin{center}
\advance\rightskip -2.9cm
\vskip-3.5cm
\includegraphics[width=0.65\textwidth]{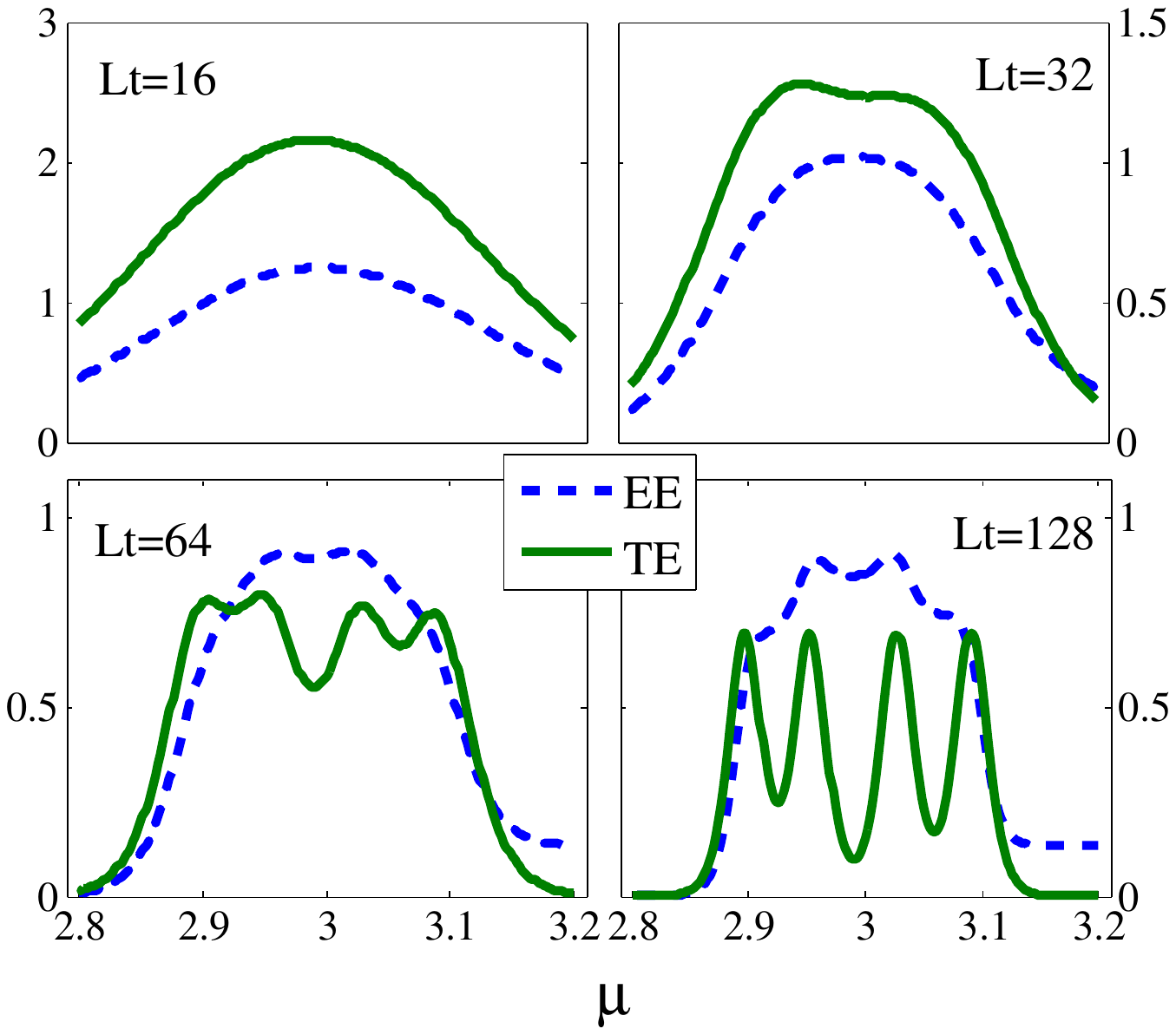}
\vskip-4.9cm
\caption{\label{fig:net} Entanglement entropy (EE, dash line)  and \TE (TE, solid line) for $\beta=0.1$, $L_x=4$ and $L_t=16$, 32, 64 and 128. 
}
\end{center}
\end{figure}

In the first SF region, we can tune $\mu$ in such way that most of the configurations will have a number of particle $n\leq L_x$. If we divide the system in two parts $A$ and $B$ of equal size $L_x/2$ we have $n+1$ ways to arrange $n_A+n_B=n$.  
The thermal density matrix 
for the whole 
system $AB$ is $
\hat{\rho}_{AB}\equiv\mathbb{T}^{L_t}/Z\ .$
If the largest eigenvalue of the transfer matrix is non degenerate with an eigenstate denoted $\ket{\Omega}$, we have 
the pure state limit
 $\lim_{L_t \rightarrow \infty} \hat{\rho}_{AB} =\ket{\Omega} \bra{\Omega}\ .$ 
We will work at finite $L_t$ and will deal with the entanglement of thermal states. In general, the eigenvalue spectrum $\{\rho_{{AB_i}}\}$ of  $\hat{\rho}_{AB}$ can then be used to define the thermal entropy
$S_T=-\sum_i \rho_{AB_i} \ln(\rho_{AB_i}).$
In order to define the entanglement entropy, 
we use the reduced density matrix $\hat{\rho}_A$ as 
$\hat{\rho}_A\equiv {\rm Tr}_B \hat{\rho}_{AB}. $
The eigenvalue spectrum $\{\rho_{A_i}\}$ of the reduced density matrix can then be used to calculate the \EE 
\begin{equation}
S_E=-\sum_i \rho_{A_i }\ln(\rho_{A_i}).\label{Eq:entropy}
\end{equation}
We use blocking methods until $A$ and $B$ are each reduced to a single site. Numerical results for $L_x=4$ are shown in Fig. \ref{fig:net} for various $L_t$. 
As $L_t$ increases, both entropies develop the distinct features stated in the introduction. 
It is easy to see that near half-filling ($n=L_x/2$), the number of ways to arrange the particles in $A$ and $B$ is of order $L_x/2$ and the entanglement 
entropy of order $\ln(L_x)$. Assuming a conformal behavior, it is possible to use this observation to measure the central charge \cite{Calabrese:2004eu}. 
$\ $
\vskip-1.3cm
$\  $
\section{Optimization through symmetry}
\begin{figure}
           \vskip-2cm
        \includegraphics[width=0.5\textwidth]{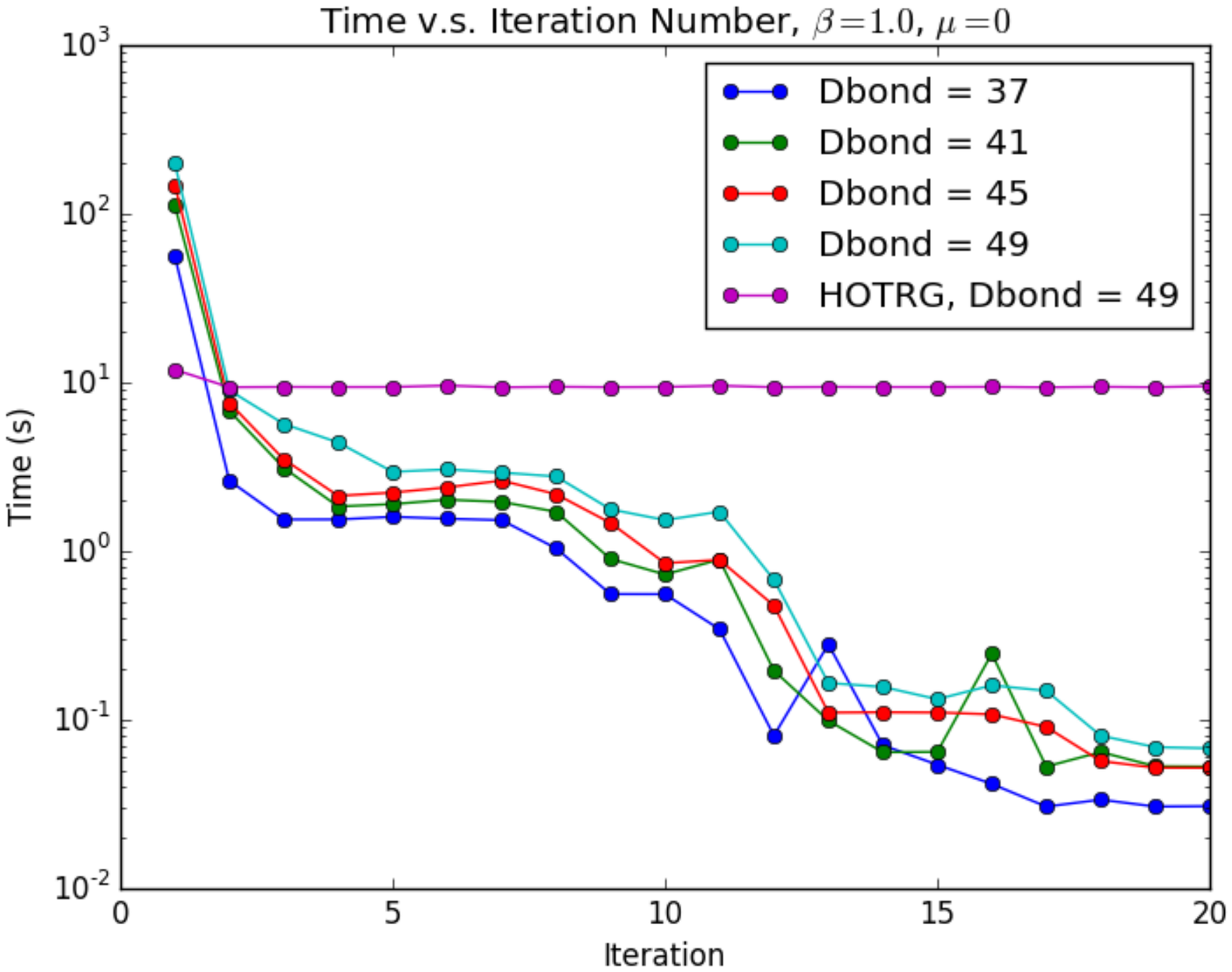}
        \includegraphics[width=0.5\textwidth]{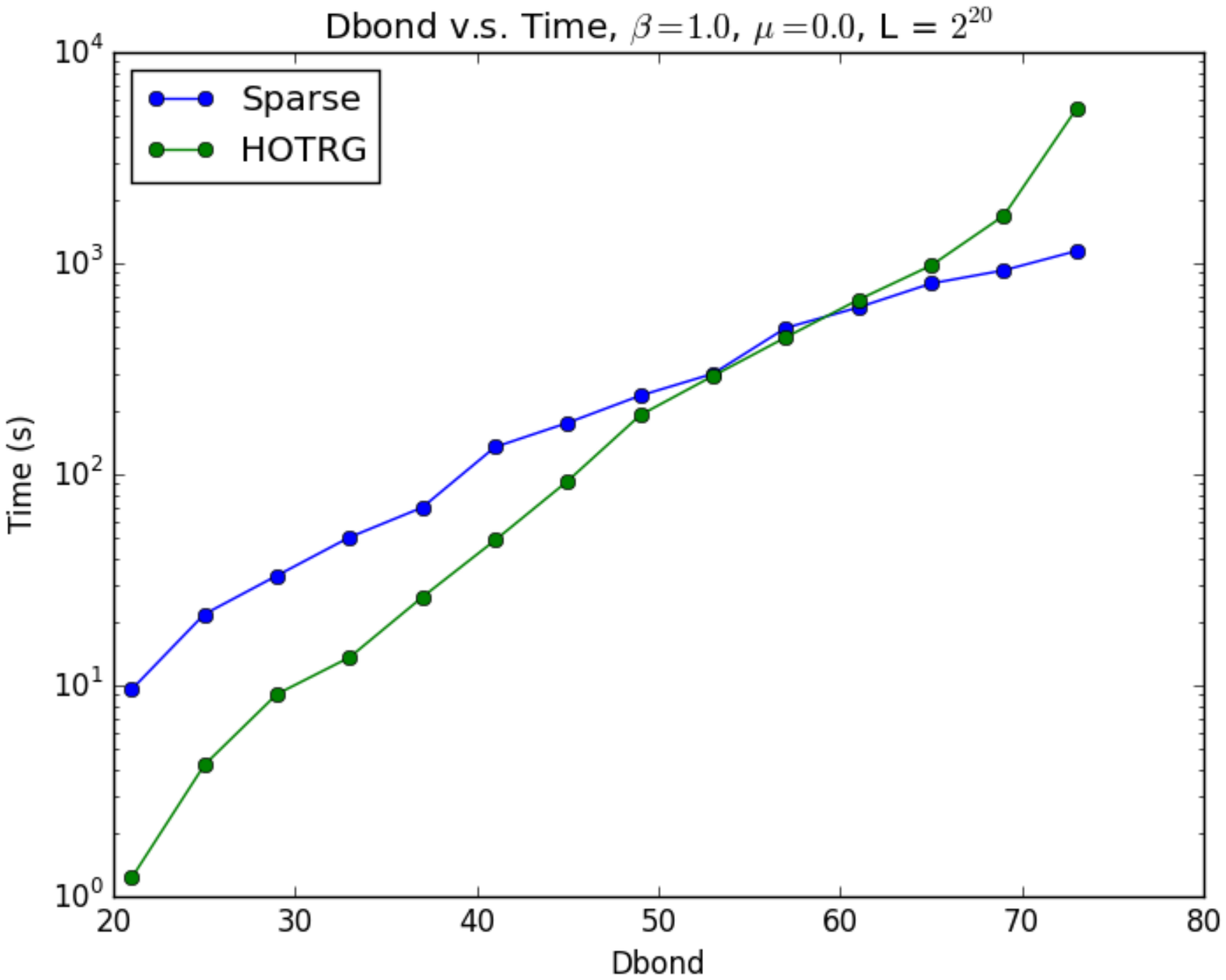}
         \vskip-2cm
        \caption{\label{fig:opt}  Computing time versus iteration number (left) and number of states (denoted $D_{bond}$, right), for the optimized method compared to the standard one (HOTRG).}
 \end{figure}

The initial $O(2)$ tensor has a built-in charge conservation . This local constraint allows a simplified parameterization
          of the tensor. For every tensor involved in contraction, one less sum/loop
          needs to be performed. In general, after one iteration, many states possess the same
          charge and a state can be labeled by its charge and a degeneracy index. We see that the charges form a distribution, then we can loop over the charges
          and treat the degeneracy index exactly as tensor indices. This gives good results when the number of charges has been reduced by iteration. The initial iteration is the worst since there is only one state per charge. However, after the first couple iterations, the iteration time decreases drastically with
          the optimized method (see Fig. \ref{fig:opt} left). After many iterations (large volume), the additional iteration time is negligible. As the number of states increases, the optimized method 
          preforms much better (see Fig. \ref{fig:opt} right). After the conference, an even better scaling was obtained. 
 \section{Conclusions}
We discussed  
the classical $O(2)$ model in 1+1 dimensions with a chemical potential $\mu$. 
The TRG formulation  allows reliable calculations of the phase diagram 
and particle density which  agree well with the worm algorithm. 
The TRG method was then used to calculate the thermal entropy and the entanglement entropy without using the replica trick. 
 At sufficiently large $L_t$, the thermal entropy and the entanglement entropy show a rich fine structure as a function of the chemical potential. 
An approximate picture of weakly interacting loops winding around the Euclidean time direction and carrying particle number one provides a particle-hole symmetry picture which justifies the mirror symmetry of the entanglement entropy with respect to half-filling. A more detailed understanding of the correspondence with the spin-1/2 XY model and calculations of the central charge are in progress \cite{liping}. 
Taking into account the particle number conservation allows more efficient TRG algorithms. A more detailed report on this question should follow \cite{judah}. It has been almost half a century since the 
important paper of Leo Kadanoff \cite{Kadanoff:1966wm} has been published and we expect that its ramifications will keep growing for many more years. 
\section{Acknowledgments}

This research was supported in part  by the Department of Energy
under Award Number DOE grant DE-SC0010114, and by the Army Research Office of the Department of Defense under Award Number W911NF-13-1-0119. 
This work utilized the Janus supercomputer, which is supported by the National Science Foundation (award number CNS-0821794) and the University of Colorado Boulder. 
The work was supported by the Natural Science Foundation of China for the Youth (Grants No.11304404). 
Y. M.  thanks the late Leo Kadanoff for enlightening discussions on TRG truncation questions.  J. U.-Y. was supported by  the Argonne Leadership Computing Facility, which is a DOE Office of Science User Facility supported under Contract DE-AC02-06CH11357, during the summers 2014 and 2015 and used computational resources while there. 

\providecommand{\href}[2]{#2}\begingroup\raggedright\endgroup


\begin{thebibliography}{10}

\bibitem{Kadanoff:1966wm}
L.~P. Kadanoff, {\it {Scaling laws for Ising models near $T_c$}},  {\em Physics}
  {\bf 2} (1966) 263--272.

\bibitem{PhysRevB.86.045139}
Z.~Y. Xie, J.~Chen, M.~P. Qin, J.~W. Zhu, L.~P. Yang, and T.~Xiang, 
{\em Phys. Rev. B} {\bf 86} (Jul, 2012) 045139, [\href{http://arxiv.org/abs/1201.1144}{{\tt arXiv:1201.1144}}].

\bibitem{prb87}
Y.~Meurice, 
  {\em Phys. Rev. B} {\bf 87} (2013), 064422,
  [\href{http://arxiv.org/abs/1211.3675}{{\tt arXiv:1211.3675}}].

\bibitem{efratirmp}
E.~Efrati, Z.~Wang, A.~Kolan, and L.~P. Kadanoff, 
{\em Rev. Mod. Phys.} {\bf 86}
  (2014) 647,  [\href{http://arxiv.org/abs/1301.6323}{{\tt arXiv:1301.6323}}].

\bibitem{Exactblocking13prd}
Y.~Liu, Y.~Meurice, M.~P. Qin, J.~Unmuth-Yockey, T.~Xiang, Z.~Y. Xie, J.~F. Yu,
  and H.~Zou, 
  {\em Phys. Rev. D} {\bf 88} (2013) 056005,  [\href{http://arxiv.org/abs/1307.6543}{{\tt arXiv:1307.6543}}].

\bibitem{prd89}
A.~Denbleyker, Y.~Liu, Y.~Meurice, M.~P. Qin, T.~Xiang, Z.~Y. Xie, J.~F. Yu,
  and H.~Zou, 
  {\em Phys. Rev. D} {\bf 89} (2014), 016008,
  [\href{http://arxiv.org/abs/1309.6623}{{\tt arXiv:1309.6623}}].

\bibitem{Shimizu:2014uva}
Y.~Shimizu and Y.~Kuramashi, 
{\em Phys. Rev. D} {\bf 90}
  (2014), 014508, [\href{http://arxiv.org/abs/1403.0642}{{\tt
  arXiv:1403.0642}}].

\bibitem{Takeda:2014vwa}
S.~Takeda and Y.~Yoshimura, 
  {\em PTEP} {\bf 2015} (2015), 043B01,
  [\href{http://arxiv.org/abs/1412.7855}{{\tt arXiv:1412.7855}}].

\bibitem{PhysRevA.90.063603}
H.~Zou, Y.~Liu, C.-Y. Lai, J.~Unmuth-Yockey, L.-P. Yang, A.~Bazavov, Z.~Y. Xie,
  T.~Xiang, S.~Chandrasekharan, S.-W. Tsai, and Y.~Meurice, 
  {\em Phys. Rev. A}
  {\bf 90} (2014) 063603,  [\href{http://arxiv.org/abs/1403.5238}{{\tt arXiv:1403.5238}}].

\bibitem{PhysRevD.92.076003}
A.~Bazavov, Y.~Meurice, S.-W. Tsai, J.~Unmuth-Yockey, and J.~Zhang, 
{\em Phys. Rev. D} {\bf 92} (2015) 076003,  [\href{http://arxiv.org/abs/1503.08354}{{\tt arXiv:1503.08354}}].

\bibitem{Saito:2014bda}
H.~Saito, M.~C. Bañuls, K.~Cichy, J.~I. Cirac, and K.~Jansen, 
{\em PoS} {\bf LATTICE2014} (2014) 302,
  [\href{http://arxiv.org/abs/1412.0596}{{\tt arXiv:1412.0596}}].

\bibitem{Buyens:2015dkc}
B.~Buyens, J.~Haegeman, F.~Verstraete, and K.~Van~Acoleyen, {\it {Tensor
  networks for gauge field theories}},  2015, these proceedings, 
\newblock \href{http://arxiv.org/abs/1511.04288}{{\tt arXiv:1511.04288}}.

\bibitem{Dittrich:2014mxa}
B.~Dittrich, S.~Mizera, and S.~Steinhaus, 
{\it {Decorated tensor network renormalization for lattice gauge theories and spin foam models}},
  \href{http://arxiv.org/abs/1409.2407}{{\tt arXiv:1409.2407}}.

\bibitem{Yang:2015rra}
L.-P. Yang, Y.~Liu, H.~Zou, Z.~Y. Xie, and Y.~Meurice, {\it {The fine structure
  of the entanglement entropy in the classical XY model}},
  \href{http://arxiv.org/abs/1507.01471}{{\tt arXiv:1507.01471}}.

\bibitem{Bruckmann:2014sla}
F.~Bruckmann and T.~Sulejmanpasic, 
{\em
  Phys. Rev. D} {\bf 90} (2014), 105010,
  [\href{http://arxiv.org/abs/1408.2229}{{\tt arXiv:1408.2229}}].

\bibitem{Bruckmann:2015sua}
F.~Bruckmann, C.~Gattringer, T.~Kloiber, and T.~Sulejmanpasic, 
 {\em Phys. Lett. B} {\bf 749} (2015) 495--501,
  [\href{http://arxiv.org/abs/1507.04253}{{\tt arXiv:1507.04253}}].

\bibitem{Kawauchi:2015heu}
H.~Kawauchi and S.~Takeda, {\it {Tensor renormalization group analysis of ${\rm
  CP}(N-1)$ model in two dimensions}},  these Proceedings, 
\newblock \href{http://arxiv.org/abs/1511.00348}{{\tt arXiv:1511.00348}}.

\bibitem{Banerjee:2010kc}
D.~Banerjee and S.~Chandrasekharan, 
 {\em Phys.Rev. D} {\bf 81} (2010) 125007,
  [\href{http://arxiv.org/abs/1001.3648}{{\tt arXiv:1001.3648}}].

\bibitem{PhysRevLett.87.160601}
N.~Prokof'ev and B.~Svistunov, 
  {\em Phys. Rev. Lett.} {\bf 87} (2001) 160601.

\bibitem{Meurice:2014tca}
Y.~Meurice, Y.~Liu, J.~Unmuth-Yockey, L.-P. Yang, and H.~Zou, {\it {Sampling
  versus Blocking}},  {\em PoS} {\bf LATTICE2014} (2014) 319,
  [\href{http://arxiv.org/abs/1411.3392}{{\tt arXiv:1411.3392}}].

\bibitem{Unmuth-Yockey:2014afa}
J.~F. Unmuth-Yockey, Y.~Meurice, J.~Osborn, and H.~Zou, {\it {Tensor
  renormalization group study of the 2d O(3) model}},  {\em PoS} {\bf
  LATTICE2014} (2014) 325,  [\href{http://arxiv.org/abs/1411.4213}{{\tt arXiv:1411.4213}}].

\bibitem{RevModPhys.52.453}
R.~Savit, 
{\em Rev.
  Mod. Phys.} {\bf 52} (Apr, 1980) 453--487.

\bibitem{Calabrese:2004eu}
P.~Calabrese and J.~L. Cardy, 
  {\em J.Stat.Mech.} {\bf 0406} (2004) P06002,
  [\href{http://arxiv.org/abs/hep-th/0405152}{{\tt hep-th/0405152}}].
  \bibitem{liping}
  Li-Ping Yang et al., work in progress.
  \bibitem{judah}
Judah Unmuth-Yockey et al., work in progress.

\end{thebibliography}
\end{document}